\documentclass[onecolumn]{aastex63}
\usepackage{graphicx}	
\usepackage{amsmath}	
\usepackage{amssymb}	
\usepackage{color}
\usepackage{array}
\usepackage{bookmark}
\usepackage{subfigure}

\shorttitle{Jet precession in GRB afterglows}
\shortauthors{Huang \& Liu}

\begin{document}
\title{Energy injection driven by precessing jets in gamma-ray burst afterglows}

\correspondingauthor{Tong Liu}
\email{tongliu@xmu.edu.cn}

\author{Bao-Quan Huang}
\affiliation{Department of Astronomy, Xiamen University, Xiamen, Fujian 361005, China}

\author[0000-0001-8678-6291]{Tong Liu}
\affiliation{Department of Astronomy, Xiamen University, Xiamen, Fujian 361005, China}

\begin{abstract}
Jet precession is considered to universally exist in different-scale astronomical phenomena, including gamma-ray bursts (GRBs). For the long-lived GRB central engine, the relativistical precessing jets will periodically inject kinetic energy into the external shocks, then significantly modulate the shapes of the light curves (LCs) in GRB afterglows. In this paper, we adopt the standard external shock model to investigate the effects of jet precession on GRB X-ray afterglows in cases with different parameters, i.e., the steady or time-dependent jet powers, precession periods, precession angles, and viewing angles. In the case where the jet powers are in steady or slow decay and the jet can sweep across the line of sight, shallow decay (or plateau) segments should appear; otherwise, a giant bump will emerge in the GRB afterglow LCs. We show that jet precession is a new plausible mechanism of the energy injection in GRBs. Moreover, some observed X-ray transients without GRB associations might be powered by the precessing jets.
\end{abstract}


\section{Introduction}

The convincing evidence regarding the existence of jet precession is that S-type and Z-type jet shapes have been observed in galaxies \citep[e.g.,][]{Miley1980,Begelman1984,Lu1990,Proctor2011,Lu2005}. The well-known X-ray binary SS 433 has a 162.5-day precession period \citep[e.g.,][]{Sarazin1980,Margon1984}. Recently, the low-frequency quasiperiodic oscillations of a black hole (BH) binary MAXI J1820+070 in the X-ray hard state have been discovered by Insight-HXMT, which probably originate from the jet precession \citep{Ma2021}. Moreover, the quasiperiodic variability of gamma-ray bursts (GRBs), especially in the BATSE data, is generally thought to be caused by the precession of relativistic jets \citep[e.g.,][]{Blackman1996,Portegies1999}.

GRBs are the most powerful electromagnetic explosions in the Universe and can be simply classified as either short- or long-duration GRBs (SGRBs and LGRBs, respectively). It is generally believed that SGRBs and LGRBs originate from the mergers of binary compact objects \citep[see review by][]{Nakar2007} and the collapse of massive stars \citep[see review by][]{Woosley2006}, respectively. Regardless, a rapidly rotating BH surrounded by a hyperaccretion disk \citep[for reviews, see][]{Liu2017a} or a massive millisecond pulsar \citep[e.g.,][]{Duncan1992,Usov1992,Dai1998b,Zhang2001} will be born in their centers. Once the directions of the angular momenta of binary compact objects are misaligned in the mergers or the explosions of core-collapse supernovae are anisotropic, the central compact object systems will be led to precess and launch precessing jets \citep[e.g.,][]{Liu2017b}. However, the precession angles are typically shallow in these scenarios.

In addition to the quasiperiodic signals in the GRB that prompt emission and the following afterglow phases, which can be explained by the jet precession, some peculiar features of light curves (LCs) are preferred in the interpretation of jet precession \citep[e.g.,][]{Reynoso2006,Lei2007,Liu2010,Stone2013}. \citet{Liu2010} proposed jet precession driven by a neutrino-cooled accretion disk \citep[e.g.,][]{Popham1999,Liu2007} around a spinning BH. The outer disk forces the BH and the inner region to precess. We modeled both the symmetric and fast-rise-exponential-decay pulses and their spectral evolutions as well as the observed gamma-ray LCs. The gravitational waves from the precession of the BH hyperaccretion systems in GRBs have also been studied \citep[e.g.,][]{Romero2010,Sun2012,Kotake2012}. \citet{Fargion2006} used a persistent, thin precessing and spinning jet within the precessing timescale, which lasts a few hours, to explain the multi-rebrightening phenomena that emerge in the afterglow phases. \citet{Hou2014a} explained the origin and time evolution of the flares in an ultra-LGRB (ULGRB), GRB 130925A, by the precessing Blandford-Znajek jets \citep{Blandford1977} launched by a long-lived BH hyperaccretion disk \citep[more than 10 ks, see, e.g.,][]{Liu2018} in the center of a massive collapsar. The same model was also used to account for the periodic signals $\sim 86~\rm s$ in the X-ray bump of another ULGRB, GRB 121027A \citep{Hou2014b}. Furthermore, \cite{Huang2019} studied jets with a precession timescale of $\sim 1~\rm s$ in the afterglow phase and described the afterglow emission for the structured jets.

\begin{figure}
\centering
\includegraphics[width=0.55\linewidth]{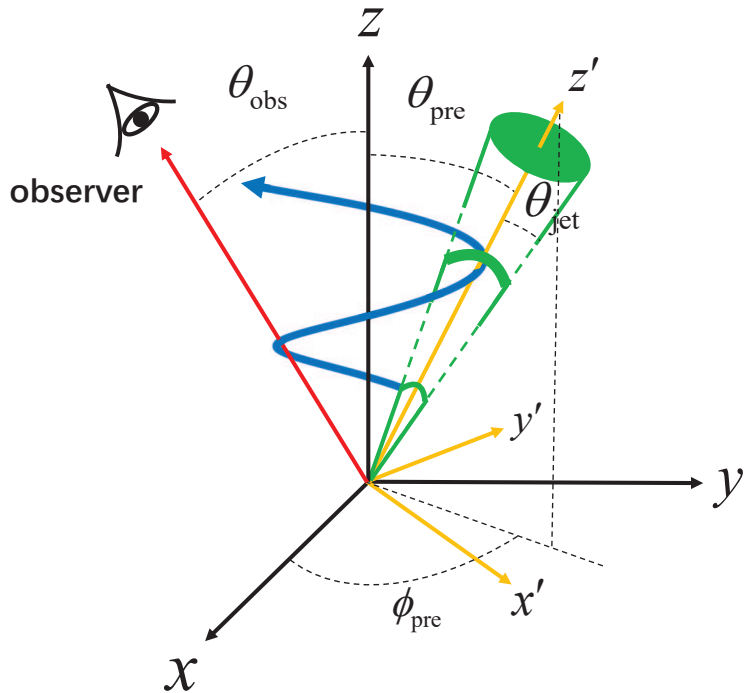}
\caption{Schematic diagram of a GRB precessing jet.}
\end{figure}

The shallow decay (or plateau) segments are generally observed in GRBs by the \emph{Swift} X-Ray Telescope \citep[XRT, see, e.g.,][]{Zhang2006,Troja2007,Rowlinson2010}. According to the following power-law decay with the indices $<-3$ and $>-3$, this decay can be divided into ``internal'' and ``ordinary'' classes, respectively \citep[e.g.,][]{Du2020}. Its timescales extend to approximately several hundred and several thousand seconds in SGRBs and LGRBs, respectively \citep[e.g.,][]{Zhang2007}. The X-ray luminosity is in the range from $\sim10^{46}$ to $10^{48} ~\rm erg~s^{-1}$ \citep[e.g.,][]{Zhao2019}. The origin of the shallow decay segment is still a mystery, and many models have been proposed in the literature, such as continuous energy injection of the spin-down luminosity of a magnetar \citep[e.g.,][]{Zhang2001,Fan2013,Du2020} or supramassive fast-rotating quark stars \citep[e.g.,][]{Li2016,Hou2018,Ouyed2020}, a jet with evolutive microphysical factors \citep[e.g.,][]{Ioka2006,Panaitescu2006} or with a bulk Lorentz factor distribution \citep[e.g.,][]{Uhm2007}, a two-component ejecta \citep[e.g.,][]{Toma2006,Yamazaki2009}, a delayed deceleration jet \citep[e.g.,][]{Duffell2015}, and a jet with a slightly off-axis view \citep[e.g.,][]{Beniamini2020}. Since the rapid decay after the internal plateaus implies an origin related to a neutron star (NS) collapsing into a BH, here, we investigate only the possible origin of the ordinary plateaus in GRB afterglows.

For a precessing jet with a shallow precession angle launched by the long-lived GRB central engine, its effects should be significantly revealed in the afterglow LCs rather than in prompt emission. In this paper, we study this energy injection mechanism to revisit the origin of the shallow decays (or plateaus) or other characteristic LCs in GRB afterglows.
This paper is organized as follows. In Section 2, we introduce the model of the GRB precessing jets and their radiation mechanism in the external shocks. The main results are given in Section 3, and conclusions and discussion are made in Section 4.

\section{Model}

\subsection{Dynamics}
To calculate the afterglow LCs of GRBs powered by a long-lived precessing jet, an approximate treatment for the precession process is adopted here. The spiral path of a precessing jet, represented by the blue curve in Figure 1, can be roughly considered to be ceaselessly filled by discrete sub-jets. The number of these discrete sub-jets is related to the jet half-opening angle $\theta_{\rm jet}$ and the precession angle $\theta_{\rm pre}$, and one should ensure that the overlaps among the discrete sub-jets are as small as possible and that the sub-jets in each period are sufficient to cover the path.

In the first precession period, due to the interaction between the sub-jets and the circumburst medium, the relativistic external forward shocks (EFSs) are generated and propagate through the medium. Because the age of the jet is much larger than the precession period, the subsequent sub-jets in each period carry out the energy injection, potentially catching up with the EFSs produced in the first period and colliding with them. The dynamics of the EFSs can be described by solving differential equations listed as follows\footnote{The dynamical descriptions of \cite{zhang2018book} adopt a proper treatment of the adiabatic loss term, which is not included in \cite{Peer2012} and \cite{huang1999}. In addition, the effect of the pressure is considered here, which is simplified by the assumption of the continuous inelastic collisions between the blast waves and the interstellar medium in \cite{huang1999}. We found that the results by using the dynamics in \citet{Peer2012} and \citet{zhang2018book} keep basically consistent, and the results by using the dynamics in \citet{huang1999} have about 1.5 times higher at the peak flux.} \citep{zhang2018book},
\begin{equation}
\frac{d\Gamma}{dt} = -\frac{\Gamma(\Gamma^{2}-1)(\hat{\gamma}\Gamma-\hat{\gamma}+1)c^{2}\frac{dm}{dt}-\Gamma(\hat{\gamma}-1)(\hat{\gamma}\Gamma^{2}-\hat{\gamma}+1)\frac{3U}{R}\frac{dR}{dt}-\Gamma^{2}
\frac{dE_{\rm inj}}{dt}}{\Gamma^{2}(M_{0}+m)c^{2}+( \hat{\gamma}^{2}\Gamma^{2}-\hat{\gamma}^{2}+3\hat{\gamma}-2)U},
\end{equation}
\begin{equation}
\frac{dU}{dt}=(1-\epsilon)(\Gamma-1)c^{2}\dfrac{dm}{dt}-
(\hat{\gamma}-1)
\left( \frac{3}{R}\frac{dR}{dt}-\dfrac{1}{\Gamma}\frac{d\Gamma}{dt} \right)U,
\end{equation}
\begin{equation}
\dfrac{dm}{dt}=4\pi R^{2}n m_{p} \dfrac{c\beta}{1-\beta},
\end{equation}
\begin{equation}
\dfrac{dR}{dt}=\dfrac{c\beta}{1-\beta},	
\end{equation}
where $\Gamma$ , $U$, $m$, and $R$ represent the bulk Lorentz factor, the internal energy, the sweep-up mass from the circumburst medium, and the distance from the central source, respectively. $E_{\rm inj}$, $M_{0}$, $\epsilon$, $n$, and $t$ are the injection energy, the initial mass, the radiation efficiency of electrons in the EFSs, the number density of the circumburst medium, and the time measured in the observer frame, respectively. Likewise, $\hat{\gamma} $ is the adiabatic index which is obtained by the formula, $\hat{\gamma}\approx(5-1.21937\xi+0.18203\xi^{2}-0.96583\xi^{3}+2.32513\xi^{4}-2.39332\xi^{5}+1.07136\xi^{6})/3$, with $ \xi\equiv \Theta/( 0.24+\Theta)$, $\Theta\approx \Gamma\beta (\Gamma\beta+1.07 (\Gamma\beta)^{2})/3(1+\Gamma\beta+1.07(\Gamma\beta)^{2})$, and $\beta=\sqrt{1-1/\Gamma^{2}}$ \citep[e.g.,][]{Peer2012}. In addition, the function form of $dE_{\rm inj}/dt$ can be simply described as
\begin{equation}
  {\frac{dE_{\rm inj}}{dt}}=
  \left\{
  \begin{array}{ll}
  2P_{\rm jet}/(1-\cos\theta_{\rm jet}),\quad T_{\rm start} < t < T_{\rm end},\\
  0,\quad \quad \quad \quad {\rm others},
  \end{array}
  \right.
\end{equation}
where $T_{\rm start} = ((i-1)\Delta t + j\tau)$, $T_{\rm end} = (i\Delta t + j\tau)$, $i$ $(=1,2,\ldots,k)$ represents the serial number of sub-jets within each period, $j$ $(=1,2,\ldots,t_{\rm end}/\tau-1)$ denotes the serial number of the precession period, $\tau$ is the precession period, and $\Delta t = \tau/k$. In this work, the value of the mass density of the medium is considered to be a constant (i.e., interstellar medium), and the evolution of $\theta_{\rm jet}$ is neglected because the sideways expansion is not important, as shown by numerical simulations \citep[e.g.,][]{Zhang2009,Chen2021}.

Two cases of the jet power $P_{\rm jet}$ are considered. A steady power is discussed in Case I, i.e., $P_{\rm jet}=P^{\rm 0}_{\rm jet}$. In Case II, a power law form
\begin{equation}
P_{\rm jet}(t)=P^{\rm 0}_{\rm jet}(t/t_{\rm 0})^{-\alpha}
\end{equation}
and a smooth broken power law form
\begin{equation}
P_{\rm jet}(t)=P^{\rm 0}_{\rm jet} [(t/t_{\rm 0})^{\alpha_{\rm r}s} + (t/t_{\rm b})^{\alpha_{\rm d}s}]^{-1/s}
\end{equation}
are taken into account, where $\alpha$, $\alpha_{\rm r}$, and $\alpha_{\rm d}$ are all temporal slopes, $t_{\rm b}$ is the break time, $s$ measures the sharpness of the break, $t_{\rm 0} \simeq R_{\rm 0} (1+z)/(2 \Gamma_0^2 c)$ is the starting time of the afterglows, with an initial radius $R_{\rm 0} = 10^{14} \,{\rm cm} $, an initial Lorentz factor $\Gamma_{\rm 0} = 200$, and  the redshift of the burst $z$, and $P^0_{\rm jet}$ is the jet power at $t_{0}$. Furthermore, in Case I, the value of the isotropic kinetic energy $E_{\rm k,iso}$ equals to $2P^{\rm 0}_{\rm jet} \Delta t/(1-\cos\theta_{\rm jet})$, while in Case II, it equals to $2\int_{(i-1)\Delta t}^{i\Delta t} P_{\rm jet}\, dt/(1-\cos\theta_{\rm jet})$. It should be noted that the afterglow starts from $t_0$; thus, the energy released before $t_0$ needs to be deducted in the first EFS process.

\subsection{Electron Distribution}

Following the evolution of the dynamics mentioned above, the evolution of the shock-accelerated electrons can be expressed as a function of the radius $R$, i.e.,
\begin{equation}
\frac{\partial }{{\partial R}}\left(\frac{{dN_{\rm{e}}^\prime }}{{d\gamma _{\rm{e}}^\prime }}\right) + \frac{\partial }{{\partial \gamma _{\rm{e}}^\prime }}\left(\dot \gamma _{\rm{e}}^\prime \frac{{dt'}}{{dR}}\frac{{dN_{\rm{e}}^\prime }}{{d\gamma _{\rm{e}}^\prime }}\right) = {{\hat{Q}'}_{{\rm{ISM}}}},
\end{equation}
where $dN_{\rm{e}}^\prime /d\gamma _{\rm{e}}^\prime $ is the instantaneous electron energy spectrum, $\gamma^\prime_{\rm e}$ is the Lorentz factor of the shock-accelerated electrons, $\dot \gamma _{\rm{e}}^\prime $ is the cooling rate of electrons with the Lorentz factor $\gamma^\prime_{\rm e}$, $dt'/dR$ = $1/\Gamma c$, and $\hat{Q}'_{\rm ISM}=\bar{K} \gamma_{\rm e}^{\prime-p}$, with $\bar{K}\approx 4\pi(p-1)R^2n_{\rm}\gamma_{\rm e,min}'^{p-1}$, is adopted to describe the injection behavior of newly shocked circumburst medium electrons \citep[e.g.,][]{Fan2008,Huang2020}, where $p$ ($> 2$) is the power law index and $\gamma^\prime_{\rm e,min}\leq\gamma^\prime_{\rm e}\leq\gamma^\prime_{\rm e,max}$ is adopted for $\gamma^\prime_{\rm e}$. It is note that quantities with a superscript accent sign are defined in the comoving frame of the EFSs. Since the shocked electrons and the magnetic fields share the fractions $\epsilon_{\rm e}$ and $\epsilon_{B}$ of the thermal energy density in the EFS downstream, the minimum Lorentz factor of the shock-accelerated electrons can be expressed as $\gamma_{\rm e,min}'=\epsilon_{\rm e}(\Gamma-1)(p-2) m_{\rm p}/(p-1)m_{\rm e}$, with $m_{\rm p}$ and $m_{\rm e}$ being the proton and electron masses, respectively, and the maximum Lorentz factor of electrons is $\gamma^\prime_{\rm e,max}=\sqrt{{9m_{\rm e}^{2}c^{4}}/{8B'e^3}}$, with $B^\prime=\sqrt{32 \pi \Gamma (\Gamma-1) n m_{\rm p} \epsilon_{B}c^2}$ being the magnetic field behind the EFSs and $c$ being the speed of light \citep[e.g.,][]{Kumar2012,Huang2020}.

\subsection{Synchrotron Radiation}

In order to calculate the observed flux of the synchrotron radiation from the sub-jets, each sub-jet is divided linearly into $300\times1000$ small emitters along $\theta^{'}$- and $\phi^{'}$-directions in the comoving spherical coordinate frame correlated to the comoving cartesian coordinate system $(x^{'}, y^{'}, z^{'})$, and wherein the sub-jet axis is aligned with $z^{'}$-axis as shown in Figure $1$. Due to the line-of-sight effect, the arrival time of photons from emitters with the radius $R$ deviated from the line of sight is delayed, compared to that in the line of sight. Since the delays, $t_{\rm de}=R(1-\cos\theta_{\rm em})/c$, only depend on the cosine value of the angle between the off-sight emitters and the line of sight, which is depicted as
\begin{equation}
\cos\theta_{\rm em}=\sin\theta_{\rm obs}\cos
\phi_{\rm obs}\sin\theta\cos\phi+\sin\theta_{\rm obs}\sin
\phi_{\rm obs}\sin\theta\sin\phi+\cos\theta_{\rm obs}\cos\theta,
\end{equation}
where $(\theta_{\rm obs},\phi_{\rm obs})$, and $(\theta,\phi)$ represent the coordinates of the observer and emitter in the observer frame, respectively. In this work, we set $\phi_{\rm obs}=0$ and the transition of the emitter coordinate from the comoving frame to the observer frame can be accomplished with following relations:
\begin{equation}
	\sin\theta\cos\phi=\sin\theta_{\rm pre}\cos\phi_{\rm pre}\cos\theta^{'}
	+\cos\theta_{\rm pre}\cos\phi_{\rm pre}\sin\theta^{'}\cos\phi^{'}
	-\sin\phi_{\rm pre}\sin\theta^{'}\sin\phi^{'},
\end{equation}
\begin{equation}
	\sin\theta\sin\phi=\sin\theta_{\rm pre}\sin\phi_{\rm pre}\cos\theta^{'}
	+\cos\theta_{\rm pre}\sin\phi_{\rm pre}\sin\theta^{'}\cos\phi^{'}
	+\cos\phi_{\rm pre}\sin\theta^{'}\sin\phi^{'},
\end{equation}
\begin{equation}
	\cos\theta=\cos\theta_{\rm pre}\cos\theta^{'}-\sin
	\theta_{\rm pre}\sin\theta^{'}\cos\phi^{'},	
\end{equation}
where $(\theta_{\rm pre},\phi_{\rm pre})$ is the coordinate of sub-jet axis in the observer frame. Thereby the observed time of an emitter at radius $R$ is $(t+t_{\rm de})(1+z)$. In addition, the synchrotron radiation power at the frequency $\nu'$ can be calculated by the following formula \citep{Rybicki1979}:
\begin{equation}
P'_{\rm syn}({\nu}')=\frac{\sqrt{3} e^3 B'}{m_{\rm e}c^2}\int\nolimits_{\gamma_{\rm e,min}'}^{\gamma_{\rm e,max}'}\bigg(\frac{dN_{\rm e}'}{d\gamma_{\rm e}'} \bigg) F \bigg(\frac{\nu'}{\nu_{\rm c}'} \bigg) d\gamma_{\rm e}',
\end{equation}
where $\nu_{\rm c}'=3 e B' \gamma_{\rm e}'^{2}/4 \pi m_{\rm e}c$ and $F({\nu'}/{\nu_{\rm c}'} )=(\nu'/\nu_{\rm c}') \int\nolimits_{\nu'/\nu_{\rm c}'}^{+ \infty} K_{5/3}(x) dx$, with $K_{5/3}(x)$ being a modified Bessel function of order 5/3.
Hence, one can calculate the observed flux by summing the flux received from the emitters at the same observer  time in a sub-jet, and the observed flux density can be derived as \citep[e.g.,][]{Granot1999}
\begin{equation}
	S_{\nu_{\rm obs}}=\frac{1+z}{4\pi D_{\rm L}^2}
	{\int}\kern-15pt\int\limits_{(\rm EATS)}{P'_{\rm syn}(\nu '){D^3}d\Omega},
\end{equation}
where ``EATS'' is the equal-arrival time surface corresponding to the same observer time in a sub-jet, $\nu^{\prime}=(1+z)\nu_{\rm obs}/D$ with $D = 1/\Gamma(1-\beta\cos\theta_{\rm em})$ being the Doppler factor of the emitters, and $D_{\rm L}$ is the luminosity distance in the standard $\Lambda$CDM cosmology model ($\Omega_M=0.27$, $\Omega_\Lambda=0.73$, and $H_0=71~\rm km~s^{-1}~Mpc^{-1}$). Further, the total observed flux becomes the accumulation of the observed flux produced on each sub-jet at the same observed time.

\begin{figure*}
\centering
\includegraphics[width=0.95\linewidth]{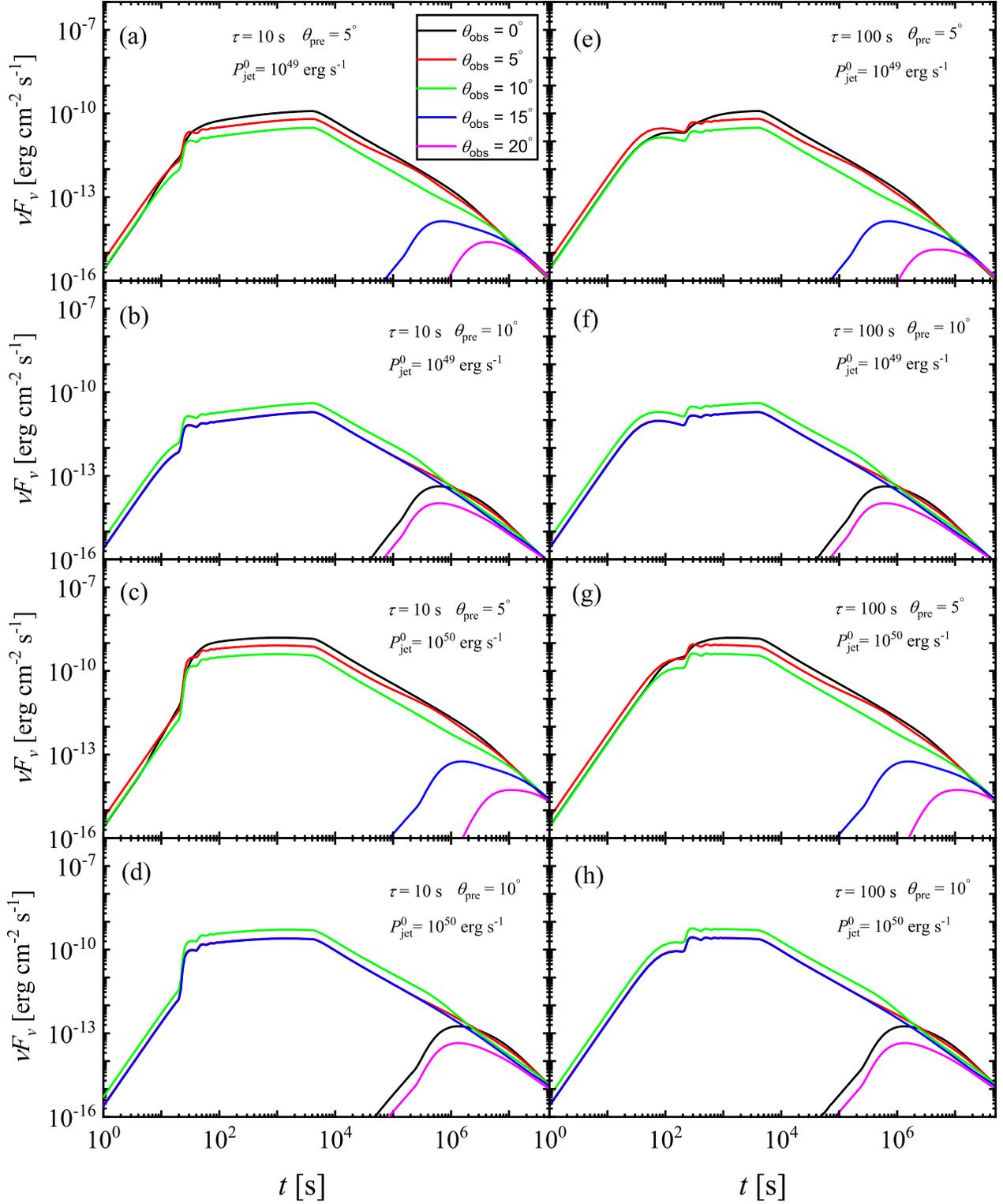}
\caption{GRB afterglow LCs of Case I in the 10 KeV band since the GRB was triggered, with the precession periods $\tau$ = 10 and 100 $\rm s$, the precession angles $\theta_{\rm pre}$ = $5^{\circ}$ and $10^{\circ}$, and the initial jet powers $P^0_{\rm jet}$ = $10^{49}$ and $10^{50} ~{\rm erg\,s^{-1}}$. The black, red, green, blue, and magenta lines denote the results for $\theta_{\rm obs}$ = $0^{\circ}$, $5^{\circ}$, $10^{\circ}$, $15^{\circ}$ and $20^{\circ}$, respectively.}
\end{figure*}

\section{Results}

\begin{figure*}
\centering
\includegraphics[width=0.95\linewidth]{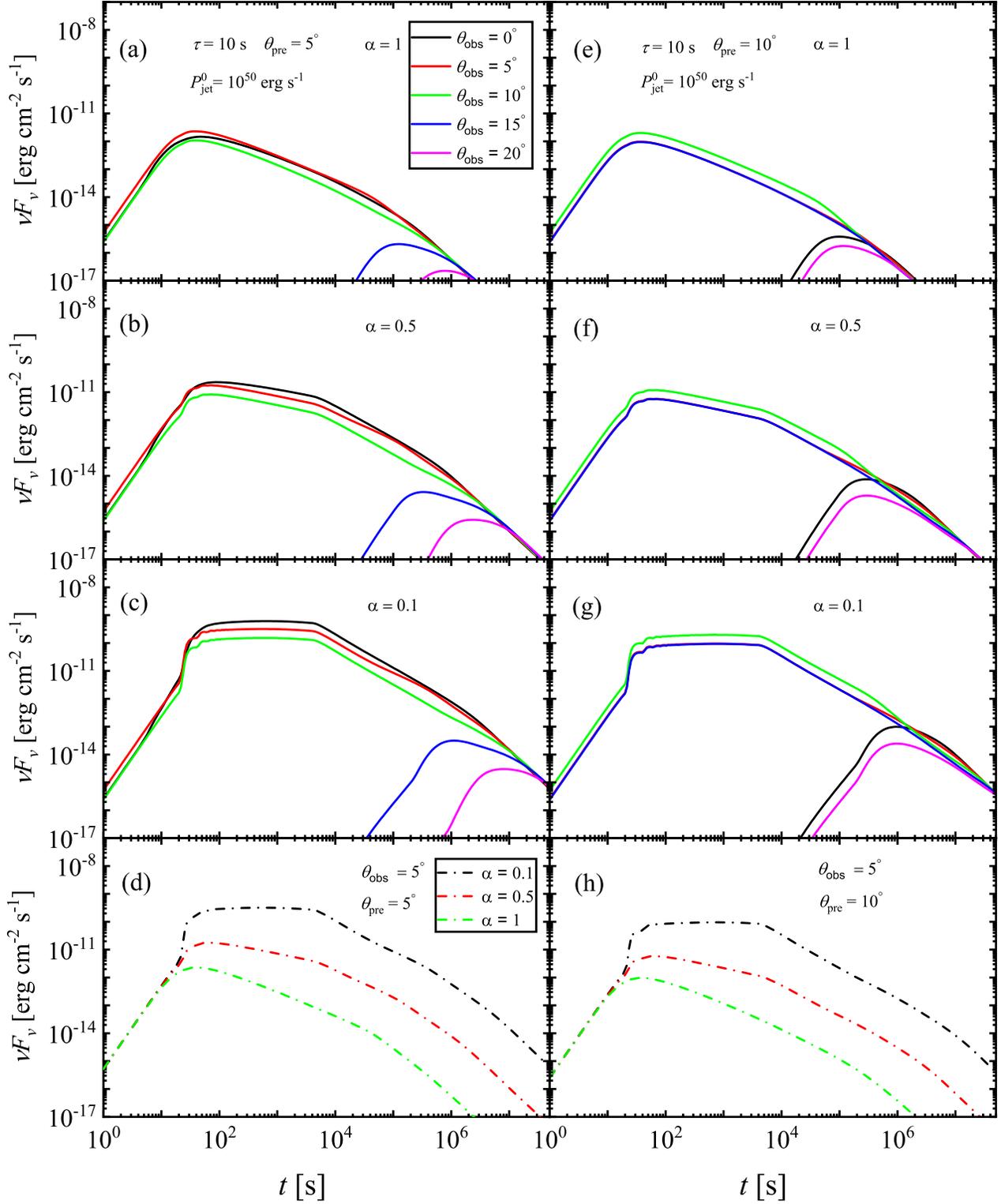}
\caption{GRB afterglow LCs of Case II with $\tau=10 \,{\rm s}$, $t_{\rm end}=2 \,{\rm ks}$, and $P_{\rm jet}^{0}=10^{50} \,{\rm erg\,s^{-1}}$. Panels (a), (b), and (c) are the results corresponding to the different $\theta_{\rm obs}$ with $\theta_{\rm pre}=5^{\circ}$ and a temporal slope $\alpha$ = 1, 0.5, and 0.1, respectively; panels (e), (f), and (g) are same, except $\theta_{\rm pre}=10^{\circ}$. Panel (d) describes the effects of the temporal slope $\alpha$ on LCs with $\theta_{\rm obs} = 5^{\circ}$ and $\theta_{\rm pre}=5^{\circ}$; panel (h) is the same, except $\theta_{\rm pre}=10^{\circ}$.}
\end{figure*}

\begin{figure*}
\centering
\includegraphics[width=0.95\linewidth]{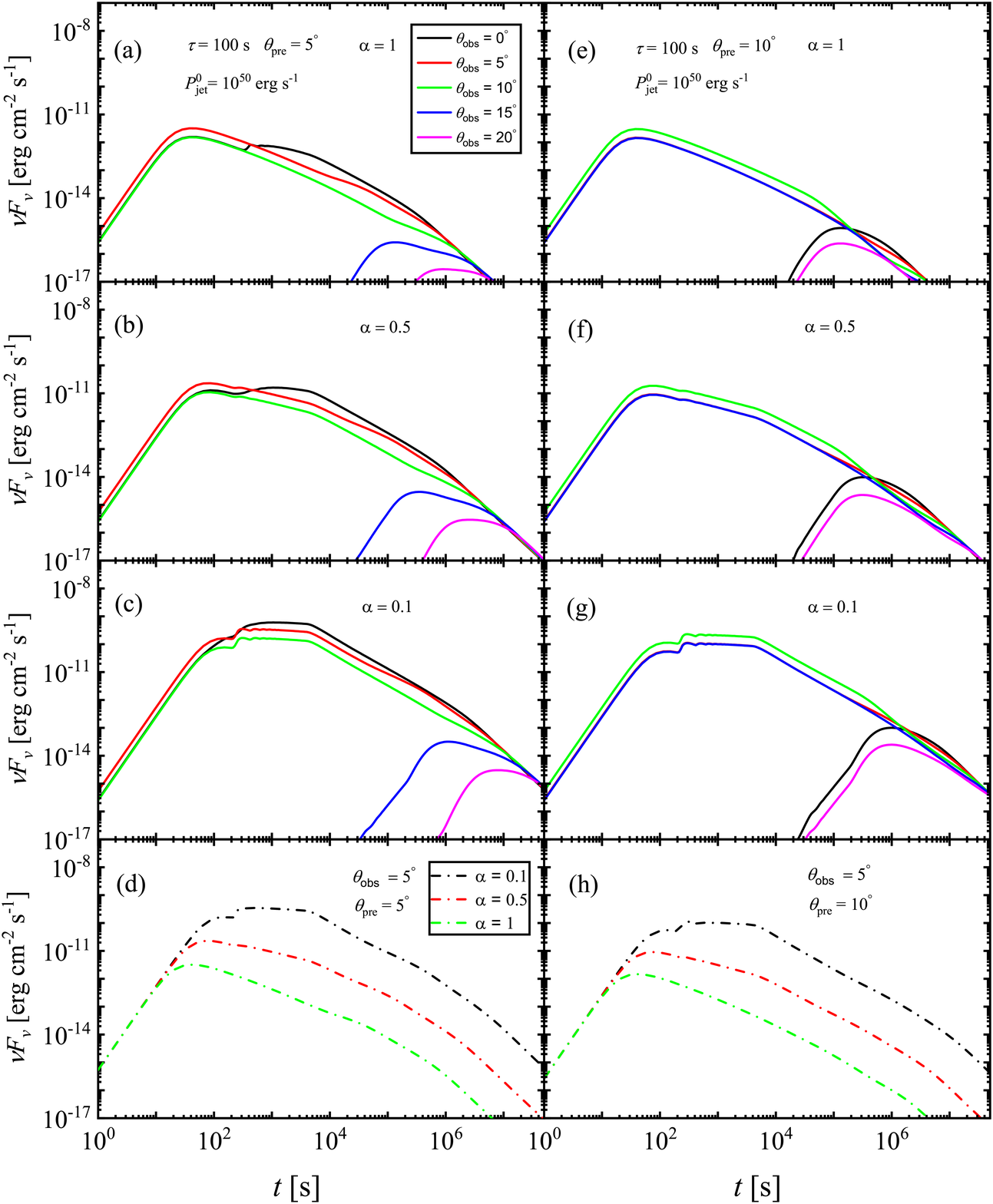}
\caption{Same as Figure 3, except $\tau=100 \,{\rm s}$.}
\end{figure*}

We calculate the X-ray afterglow LCs with the observed energy in the 10 keV band by following the above model. The typical parameter values of the EFS are set, including $\epsilon_{\rm e}= 10^{-1}$, $\epsilon_{B}=10^{-5}$, $n= 1 \,{\rm cm^{-3}}$, $p=2.3$, and $\theta_{\rm jet}=5^{\circ}$. \citet{Stone2013} obtained the expected value of the precession angle, $\sim 10^{\circ}$, under the merger of a BH and an NS; thus, we set the precession angles $\theta_{\rm pre}$ = $5^{\circ}$ and $10^{\circ}$ here. Moreover, different values of the initial jet power $P^{\rm 0}_{\rm jet}$ = $10^{49}$ and $10^{50} \,{\rm erg \, s^{-1}}$ and the precession period $\tau$ = 10 and 100 ${\rm s}$ are considered. The age of the jet, i.e., $t_{\rm end}$, is set as 2 ks, which depends on the activity of the GRB central engine and corresponds to the typical timescale of the shallow decays (or plateaus) in LGRBs. The typical redshift $z$ is set as 1.

\subsection{Case I}

\begin{figure}
\centering
\includegraphics[width=0.65\linewidth]{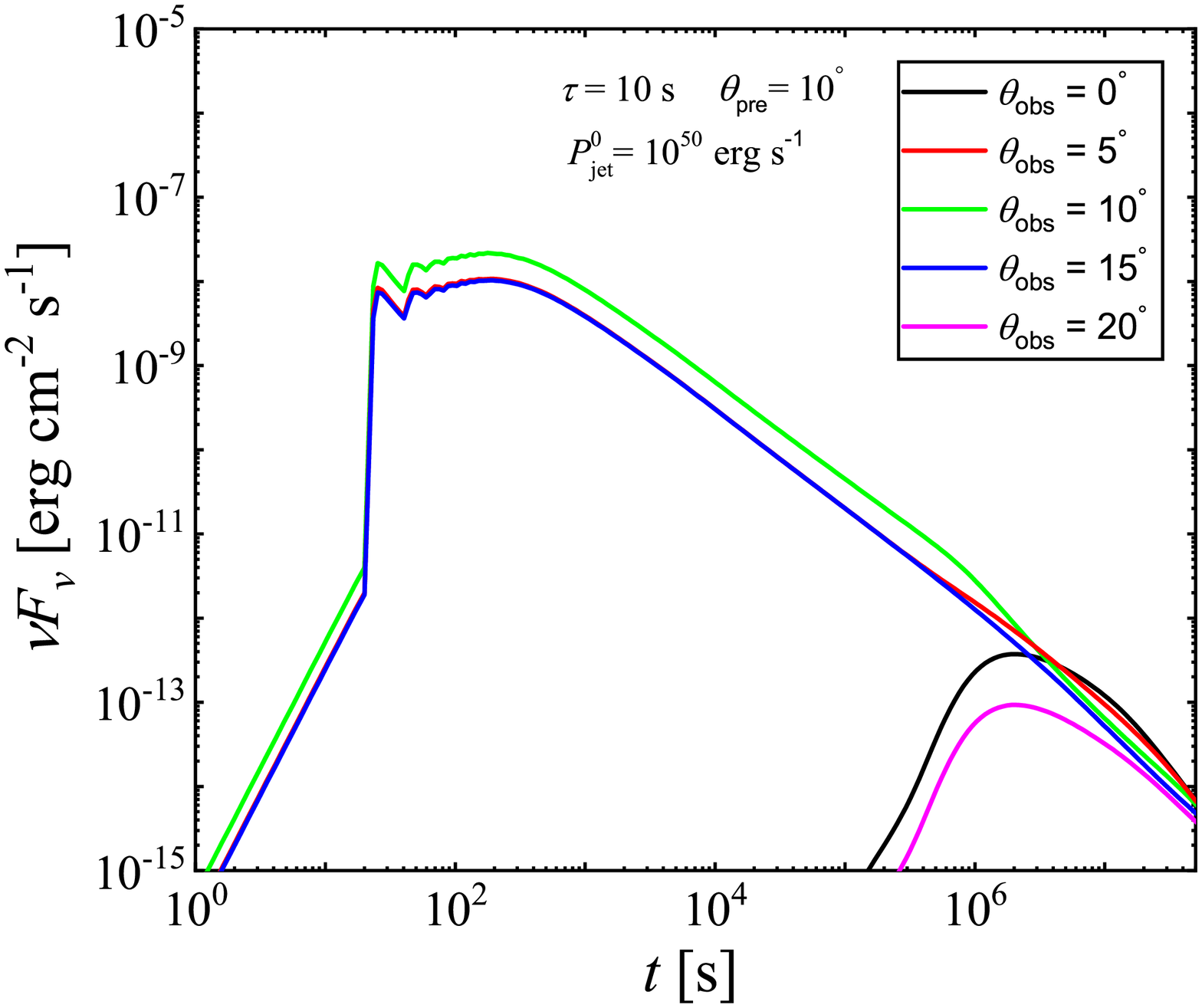}
\caption{GRB afterglow LCs of Case II with $\tau=10 \,{\rm s}$, $t_{\rm end}=2 \,{\rm ks}$, $\theta_{\rm pre}=10^{\circ}$, and $P_{\rm jet}^{0}=10^{50} \,{\rm erg\,s^{-1}}$ under the smooth broken power law evolution of jet power with $\alpha_{\rm r} = -1/2$, $\alpha_{\rm d} = 5/3$, $s=1$, and $t_{\rm b} = 1 \,{\rm ks}$.}
\end{figure}

Figure 2 shows the GRB afterglow LCs of Case I in the 10 keV band since triggering of the GRB for the precession periods $\tau$ = 10 and 100 $\rm s$, the precession angles $\theta_{\rm pre}$ = $5^{\circ}$ and $10^{\circ}$, and the initial jet powers $P_{\rm jet}$ = $10^{49}$ and $10^{50} ~{\rm erg\,s^{-1}}$. The black, red, green, blue, and magenta lines in every panel denote the results for $\theta_{\rm obs}$ = $0^{\circ}$, $5^{\circ}$, $10^{\circ}$, $15^{\circ}$, and $20^{\circ}$, respectively. In this case, one can find that the shallow decay or plateau segments (hereafter called plateaus) appear in some LCs. Some of the plateaus have tiny positive slopes.

The plateaus under the condition of a short precession period in the panels in the left column appear earlier than those in the right column, and the values of the fluxes with $P_{\rm jet} = 10^{49}\,{\rm erg\,s^{-1}}$ are definitely an order of magnitude lower than those with $P_{\rm jet} = 10^{50}\,{\rm erg\,s^{-1}}$. More importantly, when the values of the precession angles are near those of the observer angles (i.e., the jets sweep across the line of sight), the flux is higher. By contrast, the flux is lower and no plateau structure (only a bump shape) emerges when the difference between these two angles is larger, exceeding the range of the jet opening angle. Moreover, the oscillations appearing in the early stage of the plateaus might be to great extent caused by the discrete approximations of the jet precession.

\subsection{Case II}

Figure 3 shows the LCs of GRB afterglows under the power law evolution of the jet power shown in Equation (6), with $\tau=10 \,{\rm s}$, $t_{\rm end}=2 \,{\rm ks}$, and $P_{\rm jet}^{0}=10^{50} \,{\rm erg\,s^{-1}}$. Panels (a), (b), and (c) are the results with $\theta_{\rm pre}=5^{\circ}$ for temporal slopes $\alpha$ of 1, 0.5, and 0.1, respectively, and panels (e), (f), and (g) are the same as panels (a), (b), and (c), respectively, except $\theta_{\rm pre}=10^{\circ}$. The black, red, and green lines in panel (d) are the same as those of the LCs with $\theta_{\rm obs} = 5^{\circ}$ in panels (a), (b), and (c), and the lines in panel (h) are the same as those of the LCs with $\theta_{\rm obs} = 5^{\circ}$ in panels (e), (f), and (g), respectively. In this case, one can find that regardless whether $\theta_{\rm pre}=5^{\circ}$ or $10^{\circ}$, once the values of $\alpha$ are small enough, e.g., panels (c) and (g), the plateaus very clearly emerge. The slope of the plateau decreases quickly with increasing $\alpha$ until the giant bump replaces the plateau when $\alpha = 1$, as shown in panels (a) and (e). In addition, comparing panel (d) with (h), one can see that although the precessing angle is changed from $5^{\circ}$ to $10^{\circ}$, a similar plateau still exists in the LCs under the same viewing angle $\theta_{\rm obs}=5^{\circ}$, which implies that the evolution of the precessing angle does not significantly affect the plateau structure of the LCs unless the angle between the line of sight and the axis of the first sub-jet is greater than the opening angle of the sub-jets. Figure 4 is the same as Figure 3, except $\tau=100 \,{\rm s}$. Despite the difference in the precessing periods, one can see that the results implicated in Figure 4 are similar to those in Figure 3.

Figure 5 shows the LCs of GRB afterglows with $\tau=10 \,{\rm s}$, $t_{\rm end}=2 \,{\rm ks}$, and $\theta_{\rm pre}=10^{\circ}$ under a smooth broken power law evolution of the jet power, as shown in Equation (7), with $P_{\rm jet}^{0}=10^{50} \,{\rm erg\,s^{-1}}$, where the parameter values of the smooth broken power law function are $\alpha_{\rm r} = -1/2$, $\alpha_{\rm d} = 5/3$, $s=1$, and $t_{\rm b} = 1 \,{\rm ks}$. The black, red, green, blue, and magenta lines denote the results for $\theta_{\rm obs}$ = $0^{\circ}$, $5^{\circ}$, $10^{\circ}$, $15^{\circ}$, and $20^{\circ}$, respectively. One can see that a plateau appears in the LC when the viewing angle $5^{\circ} \leq \theta_{\rm obs} \leq 15^{\circ}$, but due to the quick decrease in the broken power-law jet power after $1 \,{\rm ks}$, the duration of this plateau is shorter than those in the cases with the steady or slow decay ($\alpha < 0.5$) jet powers.

From the above figures, we note that the precession period does not impact the LC shape; instead, the appearance of the plateaus nearly depends on the time evolution of the jet powers and the relations among the viewing angles, the jet half-opening angles, and the precession angles.

No matter which cases of the jet precession substantially increase the size of jet head caused by the continuous sub-jets along the spiral route to power the observable energy injections in GRB afterglows even the structured jets considered. Furthermore, the jet precession model also can apply perfectly to understand the spectral evolution of the afterglow LCs.

\section{Conclusions and discussion}

Regardless whether originating from massive collapsars or compact object mergers, jet precession might be universal in GRBs. The signature of jet precession could be imprinted in GRB observations.

In this work, we focus on GRB X-ray afterglows in the context of long-lived precessing jets. From the results, we find that a plateau or a giant bump can appear in the X-ray afterglow LCs due to the periodic energy of the jets injected into the EFSs and that the shapes of the LCs are related to the time-dependent jet power regardless whether the difference between the viewing angle and the precession angle is within the range of the jet opening angle. These results indicate that long-lived jet precession provides a new mechanism of energy injection in GRB afterglows.

In addition to the X-ray afterglow LCs, the optical or radio afterglow LCs \citep[e.g.,][]{Paradijs2000,Roming2006,Liang2013,Yi2020} should be modulated and the flux should be significantly enhanced when a larger solid angle of the sky is covered as a result of jet precession. Moreover, once a certain quasiperiodic behavior and a plateau emerge in a GRB, the information of the jet precession might be constrained.

\citet{Zhang2001} studied the effect of a continuously injected energy from a highly magnetized millisecond pulsar on GRB afterglows, and then predicted that a distinctive achromatic bump feature with the onset and duration ranging from minutes to months is presented in the afterglow LCs. In their model, the engine emits both an initial impulsive energy input $E_{\rm imp}$ as well as a continuous luminosity $L$ and the total energy of the external shock is the sum of these two parts. So, which of these two is dominant at a particular observation time $T$ depends on the values of $L$ and $E_{\rm imp}$, i.e., $T \sim E_{\rm imp}/L$ \citep{Dai1998a,Dai1998b}. In this work, we just investigate the energy injection driven by precessing jets and ignore the effects of the blast wave driven by the initial fireball that powered the prompt emission. The timescales of the plateaus in Cases I and II are about several to 10 ks, which could be longer than the typical $T$, the plateaus might be observable at least in the later stage. For instance, given $P_{\rm jet} \sim 10^{49} \,{\rm erg~s^{-1}}$ and $E_{\rm imp} \sim 10^{53} \,{\rm erg}$, if the value of the flux can meet about $10^{-11}~\rm erg ~cm^{-2}~s^{-1}$ as shown in Figure 2, the isotropic ``effective'' injection luminosity $L_{\rm eff}$ achieves about $10^{50}~\rm erg~s^{-1}$ in the precession model, thus $T \sim E_{\rm imp}/L_{\rm eff} \sim 1 ~{\rm ks}$. $L_{\rm eff}$ has almost the same value with the isotropic injection luminosity of a line-of-sight jet without precession \citep[e.g.,][]{Fan2006,Zhong2016}, and it is only a fraction of the isotropic injected power, $2P_{\rm jet}/(1-\cos\theta_{\rm jet}) \sim 10^{51}~\rm erg ~s^{-1}$. This fraction can also be roughly estimated by the proportion of the timescale of the observable sub-jets in a whole period.

In the case of BH hyperaccretion, the angular momenta between the BH and the disk will be redistributed, and massive collapsars will undergo an initially drastic accretion process lasting tens of seconds \citep[e.g.,][]{Liu2018}, which is much longer and more severe than in mergers \citep[e.g.,][]{Song2018}. Then, following the rapid decay of the accretion rate, the redistribution will tend to ease; hence, the precession period and angle might be in the quasi-steady state during the GRB afterglow phase. Thus, it might exist that the transients with X-ray plateaus but no associated GRBs if the initial jets cannot break out from the envelope and circumstances or out of the line of sight.

\citet{Xue2019} discovered a peculiar X-ray transient, CDF-S XT2, at $z$ = 0.738 in Chandra deep-field south survey. Its best-fit power-law slopes are $-0.14^{+0.03}_{-0.03}$ before the break at $2.3^{+0.4}_{-0.3}$ ks and $-2.16^{+0.26}_{-0.29}$ after the break in $0.5-7~\rm keV$ band. They proposed that it might be powered by a millisecond magnetar after a binary NS merger \citep[also see, e.g.,][]{Sun2019,Xiao2019,Ren2020}. This X-ray transient is presented from GRBs viewed far off-axis to the line of sight \citep[e.g.,][]{Dado2019,Sun2019}. Moreover, another transient in the survey, CDF-S XT1, has a power-law decay with a slope $-1.53\pm 0.27$ \citep{Bauer2017}. The LCs of these two sources are similar to our results. If we reasonably set the power-law index of the electrons $p \sim 4$ \citep[e.g.,][]{Sironi2014,Guo2015,Xiao2019}, it is easy to build up the LC of CDF-S XT2. Meanwhile, other parameters of the jet precession model are still in large value spaces. Thus we consider that these X-ray transients are possible origin from the precessing jets launched by the BH hyperaccretion or magnetars in the scenarios of the massive collapsars or compact object mergers.

\acknowledgments
We thank Dr. Da-Bin Lin for the helpful discussion. This work was supported by the National Natural Science Foundation of China under grant 11822304.


\begin{thebibliography}{99}
\bibitem[Bauer et al.(2017)]{Bauer2017} Bauer, F. E., Treister, E., Schawinski, K., et al. 2017, \mnras, 467, 4841
\bibitem[Begelman et al.(1984)]{Begelman1984} Begelman, M.~C., Blandford, R.~D., \& Rees, M.~J.\ 1984, Reviews of Modern Physics, 56, 255
\bibitem[Beniamini et al.(2020)]{Beniamini2020} Beniamini, P., Duque, R., Daigne, F., et al.\ 2020, \mnras, 492, 2847
\bibitem[Blackman et al.(1996)]{Blackman1996} Blackman, E.~G., Yi, I., \& Field, G.~B.\ 1996, \apjl, 473, L79
\bibitem[Blandford \& Znajek(1977)]{Blandford1977} Blandford, R.~D. \& Znajek, R.~L.\ 1977, \mnras, 179, 433
\bibitem[Chen \& Zhang(2021)]{Chen2021} Chen, L. \& Zhang, B.\ 2021, \apj, 906, 105
\bibitem[Dado \& Dar(2019)]{Dado2019} Dado, S. \& Dar, A.\ 2019, \apjl, 884, L44
\bibitem[Dai \& Lu(1998a)]{Dai1998a} Dai, Z.~G. \& Lu, T.\ 1998a, \aap, 333, L87
\bibitem[Dai \& Lu(1998b)]{Dai1998b} Dai, Z.~G., \& Lu, T.\ 1998b, \prl, 81, 4301
\bibitem[Du (2020)]{Du2020} Du, S. 2020, \apj, 901, 75
\bibitem[Duffell \& MacFadyen(2015)]{Duffell2015} Duffell, P.~C. \& MacFadyen, A.~I.\ 2015, \apj, 806, 205
\bibitem[Duncan \& Thompson(1992)]{Duncan1992} Duncan, R. C., \& Thompson, C. 1992, \apjl, 392, 9
\bibitem[Fan \& Piran(2006)]{Fan2006} Fan, Y. \& Piran, T.\ 2006, \mnras, 369, 197
\bibitem[Fan et al.(2008)]{Fan2008} Fan, Y.-Z., Piran, T., Narayan, R., et al.\ 2008, \mnras, 384, 1483
\bibitem[Fan et al.(2013)]{Fan2013} Fan, Y.-Z., Yu, Y.-W., Xu, D., et al.\ 2013, \apjl, 779, L25
\bibitem[Fargion \& Grossi(2006)]{Fargion2006} Fargion, D. \& Grossi, M.\ 2006, Chinese Journal of Astronomy and Astrophysics Supplement, 6, 342
\bibitem[Granot et al.(1999)]{Granot1999} Granot, J., Piran, T., \& Sari, R.\ 1999, \apj, 513, 679
\bibitem[Guo et al.(2015)]{Guo2015} Guo, F., Liu, Y.-H., Daughton, W., \& Li, H. 2015, \apj, 806, 167
\bibitem[Hou et al.(2014b)]{Hou2014b} Hou, S.-J., Gao, H., Liu, T., et al.\ 2014b, \mnras, 441, 2375
\bibitem[Hou et al.(2014a)]{Hou2014a} Hou, S.-J., Liu, T., Gu, W.-M., et al.\ 2014a, \apjl, 781, L19
\bibitem[Hou et al.(2018)]{Hou2018} Hou, S.-J., Liu, T., Xu, R.-X., et al.\ 2018, \apj, 854, 104
\bibitem[Huang et al.(2019)]{Huang2019} Huang, B.-Q., Lin, D.-B., Liu, T., et al.\ 2019, \mnras, 487, 3214
\bibitem[Huang et al.(2020)]{Huang2020} Huang, B.-Q., Liu, T., Huang, F., et al.\ 2020, \apj, 904, 17
\bibitem[Huang et al.(1999)]{huang1999} Huang, Y.~F., Dai, Z.~G., \& Lu, T.\ 1999, \mnras, 309, 513
\bibitem[Ioka et al.(2006)]{Ioka2006} Ioka, K., Toma, K., Yamazaki, R., \& Nakamura, T. 2006, \aap, 458, 7
\bibitem[Kotake et al.(2012)]{Kotake2012} Kotake, K., Takiwaki, T., \& Harikae, S.\ 2012, \apj, 755, 84
\bibitem[Kumar et al.(2012)]{Kumar2012} Kumar, P., Hern{\'a}ndez, R.~A., Bo{\v{s}}njak, {\v{Z}}., et al.\ 2012, \mnras, 427, L40
\bibitem[Lei et al.(2007)]{Lei2007} Lei, W.~H., Wang, D.~X., Gong, B.~P., et al.\ 2007, \aap, 468, 563
\bibitem[Li et al.(2016)]{Li2016} Li, A., Zhang, B., Zhang, N.-B., et al.\ 2016, \prd, 94, 083010
\bibitem[Liang et al.(2013)]{Liang2013} Liang, E.-W., Li, L., Gao, H., et al.\ 2013, \apj, 774, 13
\bibitem[Liu et al.(2007)]{Liu2007} Liu, T., Gu, W.-M., Xue, L., et al.\ 2007, \apj, 661, 1025
\bibitem[Liu et al.(2017a)]{Liu2017a} Liu, T., Gu, W.-M., \& Zhang, B.\ 2017a, \nar, 79, 1
\bibitem[Liu et al.(2010)]{Liu2010} Liu, T., Liang, E.-W., Gu, W.-M., et al.\ 2010, \aap, 516, A16
\bibitem[Liu et al.(2017b)]{Liu2017b} Liu, T., Lin, C.-Y., Song, C.-Y., et al.\ 2017b, \apj, 850, 30
\bibitem[Liu et al.(2018)]{Liu2018} Liu, T., Song, C.-Y., Zhang, B., et al.\ 2018, \apj, 852, 20
\bibitem[Lu(1990)]{Lu1990} Lu, J.-F. 1990, \aap, 229, 424
\bibitem[Lu \& Zhou(2005)]{Lu2005} Lu, J.-F., \& Zhou, B.-Y. 2005, \apjl, 635, L17
\bibitem[Ma et al.(2021)]{Ma2021} Ma, X., Tao, L., Zhang, S.-N., et al.\ 2021, Nature Astronomy, 5, 94
\bibitem[Margon(1984)]{Margon1984} Margon, B.\ 1984, \araa, 22, 507
\bibitem[Miley(1980)]{Miley1980} Miley, G.\ 1980, \araa, 18, 165
\bibitem[Nakar(2007)]{Nakar2007} Nakar, E.\ 2007, \physrep, 442, 166
\bibitem[Ouyed et al.(2020)]{Ouyed2020} Ouyed, R., Leahy, D., \& Koning, N.\ 2020, Research in Astronomy and Astrophysics, 20, 027
\bibitem[Panaitescu et al.(2006)]{Panaitescu2006} Panaitescu, A., M{\'e}sz{\'a}ros, P., Burrows, D., et al.\ 2006, \mnras, 369, 2059
\bibitem[Pe'er(2012)]{Peer2012} Pe'er, A.\ 2012, \apjl, 752, L8
\bibitem[Popham et al.(1999)]{Popham1999} Popham, R., Woosley, S.~E., \& Fryer, C.\ 1999, \apj, 518, 356
\bibitem[Portegies Zwart et al.(1999)]{Portegies1999} Portegies Zwart, S.~F., Lee, C.-H., \& Lee, H.~K.\ 1999, \apj, 520, 666
\bibitem[Proctor(2011)]{Proctor2011} Proctor, D.~D.\ 2011, \apjs, 194, 31
\bibitem[Ren et al.(2020)]{Ren2020} Ren, X., Wei, D., Zhu, Z., et al.\ 2020, \aap, 641, A56
\bibitem[Reynoso et al.(2006)]{Reynoso2006} Reynoso, M.~M., Romero, G.~E., \& Sampayo, O.~A.\ 2006, \aap, 454, 11
\bibitem[Romero et al.(2010)]{Romero2010} Romero, G.~E., Reynoso, M.~M., \& Christiansen, H.~R.\ 2010, \aap, 524, A4
\bibitem[Roming et al.(2006)]{Roming2006} Roming, P.~W.~A., Schady, P., Fox, D.~B., et al.\ 2006, \apj, 652, 1416
\bibitem[Rowlinson et al.(2010)]{Rowlinson2010} Rowlinson, A., O'Brien, P. T., Tanvir, N. R., et al. 2010, \mnras, 409, 531
\bibitem[Rybicki \& Lightman(1979)]{Rybicki1979} Rybicki, G.~B., \& Lightman, A.~P.\ 1979, Radiative processes in astrophysics (New York: Interscience)
\bibitem[Sarazin et al.(1980)]{Sarazin1980} Sarazin, C.~L., Begelman, M.~C., \& Hatchett, S.~P.\ 1980, \apjl, 238, L129
\bibitem[Sironi \& Spitkovsky(2014)]{Sironi2014} Sironi, L., \& Spitkovsky, A. 2014, \apj, 783, L21
\bibitem[Song et al.(2018)]{Song2018} Song, C.-Y., Liu, T., \& Li, A.\ 2018, \mnras, 477, 2173
\bibitem[Stone et al.(2013)]{Stone2013} Stone, N., Loeb, A., \& Berger, E.\ 2013, \prd, 87, 084053
\bibitem[Sun et al.(2019)]{Sun2019} Sun, H., Li, Y., Zhang, B.-B., et al.\ 2019, \apj, 886, 129
\bibitem[Sun et al.(2012)]{Sun2012} Sun, M.-Y., Liu, T., Gu, W.-M., et al.\ 2012, \apj, 752, 31
\bibitem[Toma et al.(2006)]{Toma2006} Toma, K., Ioka, K., Yamazaki, R., \& Nakamura, T. 2006, \apjl, 640, L139
\bibitem[Troja et al.(2007)]{Troja2007} Troja, E., Cusumano, G., O'Brien, P. T., et al. 2007, \apj, 665, 599
\bibitem[Uhm \& Beloborodov(2007)]{Uhm2007} Uhm, Z.~L. \& Beloborodov, A.~M.\ 2007, \apjl, 665, L93
\bibitem[Usov(1992)]{Usov1992} Usov, V.~V.\ 1992, \nat, 357, 472
\bibitem[van Paradijs et al.(2000)]{Paradijs2000} van Paradijs, J., Kouveliotou, C., \& Wijers, R.~A.~M.~J.\ 2000, \araa, 38, 379
\bibitem[Woosley \& Bloom(2006)]{Woosley2006} Woosley, S.~E. \& Bloom, J.~S.\ 2006, \araa, 44, 507
\bibitem[Xiao et al.(2019)]{Xiao2019} Xiao, D., Zhang, B.-B., \& Dai, Z.-G.\ 2019, \apjl, 879, L7
\bibitem[Xue et al.(2019)]{Xue2019} Xue, Y.~Q., Zheng, X.~C., Li, Y., et al.\ 2019, \nat, 568, 198
\bibitem[Yamazaki(2009)]{Yamazaki2009} Yamazaki, R.\ 2009, \apjl, 690, L118
\bibitem[Yi et al.(2020)]{Yi2020} Yi, S.-X., Wu, X.-F., Zou, Y.-C., et al.\ 2020, \apj, 895, 94
\bibitem[Zhang(2007)]{Zhang2007} Zhang, B.\ 2007, Advances in Space Research, 40, 1186
\bibitem[Zhang(2018)]{zhang2018book} Zhang, B.\ 2018, The Physics of Gamma-Ray Bursts (Cambridge: Cambridge Univ. Press)
\bibitem[Zhang et al.(2006)]{Zhang2006} Zhang, B., Fan, Y.~Z., Dyks, J., et al.\ 2006, \apj, 642, 354
\bibitem[Zhang \& M{\'e}sz{\'a}ros(2001)]{Zhang2001} Zhang, B. \& M{\'e}sz{\'a}ros, P.\ 2001, \apjl, 552, L35
\bibitem[Zhang \& MacFadyen(2009)]{Zhang2009} Zhang, W. \& MacFadyen, A.\ 2009, \apj, 698, 1261
\bibitem[Zhao et al.(2019)]{Zhao2019} Zhao, L., Zhang, B., Gao, H., et al.\ 2019, \apj, 883, 97
\bibitem[Zhong et al.(2016)]{Zhong2016} Zhong, S.-Q., Xin, L.-P., Liang, E.-W., et al.\ 2016, \apj, 831, 5
\end{thebibliography}
\end{document}